\documentclass[aps,preprint,showpacs]{revtex4}
\usepackage{epsfig}
\usepackage{amsmath}
\usepackage{amssymb}
\usepackage{amsfonts}
\usepackage{enumerate}

\begin{document}
%\draft
\title{Spin relaxation of ``upstream'' electrons in quantum wires: Failure of 
the drift diffusion model}
\author{Sandipan Pramanik and Supriyo Bandyopadhyay \footnote{Corresponding 
author. E-mail:
sbandy@vcu.edu}}
\affiliation{Department of Electrical and Computer Engineering, Virginia
Commonwealth University, Richmond, VA 23284, USA}
\author{Marc Cahay}
\affiliation{Department of Electrical and Computer Engineering and Computer
Science,
University of Cincinnati, Cincinnati, OH 45221, USA}

\begin{abstract}

The classical drift diffusion (DD) model of spin transport treats spin 
relaxation via 
an empirical parameter known as the ``spin diffusion length''. According to this 
model, the ensemble averaged spin of electrons drifting and diffusing in a 
solid  decays exponentially 
with distance due to spin dephasing interactions. The characteristic length 
scale associated with this decay is the 
spin diffusion length. The DD model also predicts that this length is different 
for ``upstream'' electrons traveling in a decelerating electric field than for 
``downstream'' electrons traveling in an accelerating field. However this 
picture ignores energy quantization in 
confined systems (e.g. quantum wires) and therefore fails to capture the 
non-trivial influence of subband structure on spin relaxation. Here we 
highlight this influence by simulating upstream spin transport in a 
multi-subband quantum wire, in the presence of D'yakonov-Perel' spin relaxation, 
using a semi-classical model that accounts for the subband structure rigorously. 
 We find 
that upstream spin transport has a complex 
dynamics that defies the simplistic 
definition of a ``spin 
diffusion length''.
 In fact, spin does not decay exponentially or even monotonically with distance, 
and the drift diffusion picture fails to explain the qualitative behavior, let 
alone predict quantitative features accurately.
Unrelated to spin transport, we 
also find that upstream electrons undergo a ``population inversion'' as a 
consequence of the energy dependence of the density of states in a quasi 
one-dimensional structure.
 
\end{abstract}

\maketitle
\pagebreak

\section{Introduction}
Spin transport in semiconductor structures is a subject 
of
much
interest  from the perspective of both fundamental physics and device
applications. A number of different formalisms have been used to study this
problem, primary among which are a
classical drift
diffusion approach \cite{flatte1,flatte2,saikin1}, a kinetic theory 
approach 
\cite{wu}, and a microscopic semiclassical approach 
\cite{bournel1,bournel2,bournel3,bournel4,saikin,saikin1,pramanik_prb,pramanik_apl}. 
The central result of the drift diffusion approach is a differential equation 
that describes the 
spatial and temporal evolution of carriers with a certain spin polarization 
$n_{\sigma}$. Ref. \cite{saikin1} derived this equation for a number of special 
cases starting from the Wigner distribution function. In a coordinate system 
where the x-axis coincides with the direction of electric field driving 
transport, this equation is of the form: 
\begin{equation}
{\frac{\partial n_{\sigma}}{\partial t}} - {\bf D} {\frac{\partial^2 
n_{\sigma}}{\partial x^2}} - {\bf A} {\frac{\partial n_{\sigma}}{\partial 
x}} + {\bf B} n_{\sigma} = 0
\label{spin}
\end{equation}
where
\begin{eqnarray}
{\bf D} & = &  \left( \begin{array}{ccc}
             D & 0 & 0\\
             0 & D & 0 \\
             0 & 0 & D \\
             \end{array}   \right)  ~,\\
\end{eqnarray}
$D$ is the diffusion coefficient, and ${\bf A}$ and ${\bf B}$ are dyadics 
(9-component tensors) that depend on $D$, the mobility $\mu$ and the spin orbit 
interaction strength in the material.

Solutions of Equation (\ref{spin}), with appropriate boundary conditions, 
predict that the ensemble averaged spin
$\left|\langle \bf{S}\rangle\right|(x)$ = $\sqrt{\langle S_x\rangle^2(x) + 
\langle S_y\rangle^2(x) + \langle S_z\rangle^2(x)}$ should decay 
exponentially with $x$ according to:
\begin{equation}
\left |\langle \bf{S}\rangle\right | (x)= \left|\langle 
\bf{S}\rangle\right|(0)e^{-x/L}
\label{decay}
\end{equation}
where 
\begin{equation}
{\frac{1}{L}} = {\frac{\mu E}{2 D}} + \sqrt{\left ( {\frac{\mu E}{2 D}}\right )^2 + C^2}. 
\label{diffusion}
\end{equation}
Here $E$ is the strength of the driving electric field and $C$ is a parameter 
related to the spin orbit interaction strength.

 The quantity $L$ is the characteristic length over which $\left|\langle 
\bf{S}\rangle\right|$ decays 
to $1/e$ times its original value. Therefore, it is defined as the ``spin 
diffusion length''. Equation (\ref{diffusion}) clearly shows that spin diffusion 
length depends on the {\it sign} of the electric field $E$. It is smaller for 
upstream transport (when $E$ is positive) than for downstream transport (when 
$E$ is negative).

This difference assumes importance
in the context of spin injection from a metallic ferromagnet into a
semiconducting paramagnet. Ref. \cite{flatte1} pointed out that the 
spin injection
efficiency across the interface between these materials depends on the
difference between the quantities
$L_s/\sigma_s$  and $L_m/\sigma_m$ where $L_s
$ is the spin diffusion length in the semiconductor, $\sigma_s$ is the 
conductivity of the
semiconductor,
$\sigma_m$ is the conductivity of the metallic ferromagnet, and $L_m$ is 
the
spin diffusion length in the metallic ferromagnet. Generally, $\sigma_m >> 
\sigma_s$. However, at
sufficiently high retarding electric field, $L_s << L_m$, so that 
$L_s/\sigma_s\approx L_m/\sigma_m$. When
this equality
is established, the spin injection efficiency is maximized. Thus, ref.
\cite{flatte1}
claimed that it is possible to circumvent the infamous ``conductivity
mismatch''
problem \cite{molenkamp}, which inhibits efficient spin injection across a 
metal-semiconductor interface, by applying a
high retarding electric field in the semiconductor
close to the interface. A tunnel barrier between the
ferromagnet and semiconductor \cite{rashba_tunnel} or a Schottky barrier 
\cite{hanbicki1, hanbicki2} at the
interface does essentially this and therefore improves spin injection.

The result of ref. \cite{flatte1} depends on the validity of the drift diffusion 
model and Equation (\ref{decay}) which predicts an exponential decay of spin 
polarization in space. Without  the exponential decay, one cannot even define a 
``spin diffusion length'' $L$. The question then is whether one expects to see 
the exponential decay under all circumstances, particularly in quantum confined 
structures such as quantum wires. The answer to this question is in the 
{\it negative}.
Equation (\ref{spin}), and similar equations derived within the drift diffusion 
model, do not account for energy quantization in quantum confined systems and 
neglect the influence of subband structure on spin depolarization. This is a 
serious shortcoming since in a semiconductor quantum wire, the spin orbit 
interaction strength is {\it different} in different subbands. It is this 
difference that results in D'yakonov-Perel' (D-P) spin relaxation in quantum 
wires. Without this difference, the D-P relaxation will be completely absent in 
quantum wires and the corresponding spin diffusion length will be always 
infinite  
\cite{pramanik_transnano}. The suband structure is therefore vital to spin 
relaxation.

\section{SEMICLASSICAL MODEL OF SPIN RELAXATION}

In this paper, we have studied spin relaxation using a microscopic semiclassical 
model that is derived from the Liouville equation for the spin density matrix 
\cite{blum, privman}. Our model has been described in ref. \cite{pramanik_prb} 
and wil not be repeated here. This model allows us to study D'yakonov-Perel' 
spin relaxation taking into account the detailed subband structure in the system 
being studied.
 
 In technologically important semiconductors, such as 
GaAs, spin relaxation is dominated by the D'yakonov-Perel' (D-P) mechanism 
\cite{dp}. 
This mechanism  arises from the Dresselhaus \cite{dresselhaus} and Rashba 
\cite{rashba} spin 
orbit interactions that act as momentum dependent effective magnetic fields 
${\bf B({\bf k})}$. 
 An electron's spin polarization vector ${\bf S}$ precesses about ${\bf B ({\bf 
k})}$ according to the equation 
 \begin{equation}
{\frac{d {\bf S}}{dt}} = {\bf \Omega ({\bf k}) } \times {\bf S}
\label{precession}
\end{equation}
where $\Omega (k)$ is the angular frequency of spin precession and   is related 
to ${\bf B ({\bf k})}$ as ${\bf \Omega (k) } = 
(e/m^*) {\bf B ({\bf k})}$, where $m^*$ is the electron's effective mass. If the 
direction of {\bf B({\bf 
k})} changes randomly due to carrier scattering which changes {\bf k}, then 
ensemble averaging over the spins of a large number of electrons will lead to a 
decay of the ensemble averaged spin in space and time. 
This is the physics of 
the D-P relaxation in bulk and quantum wells.
In a quantum wire, the direction 
of 
${\bf k}$ never changes (it is always along the axis of the wire) in spite of 
scattering.
Nevertheless, there can be D-P relaxation in a {\it {multi-subband}} quantum
wire, as we explain in the next paragraphs.

We will consider a quantum wire of rectangular cross-section with its axis along 
the [100] crystallographic orientation (which we label the x-axis), and a 
symmetry breaking electric field $ E_y$ is applied along the y-axis to induce the 
Rashba interaction (refer Fig. \ref{model}). Then, the components of the vector ${\bf \Omega (k) }$ due 
to 
the Dresselhaus and Rashba interactions are given by
\begin{eqnarray}
{\bf \Omega_D (k)} & = & {\frac{2 a_{42}}{\hbar}} \left [ \left ({\frac{n
\pi}{W_y}} 
\right )^2 - \left ({\frac{m \pi}{W_z}} \right )^2 \right ] k_x \hat{x}; ~~~ 
(W_z 
> W_y)
\nonumber \\
{\bf \Omega_R (k)} & = & {\frac{2 a_{46}}{\hbar}} E_y k_x \hat{z}
\end{eqnarray}
where $a_{42}$ and $a_{46}$ are material constants, $(m,n)$ are the transverse 
subband indices, $k_x$ is the wavevector along 
the axis of the quantum wire, and $W_z, W_y$ are the 
transverse dimensions of the quantum wire along the z- and y-directions.
Therefore,
\begin{eqnarray}
{\bf B ({\bf k})} & = &  {\frac{m^*}{e}}  \left [ {\bf \Omega_D (k)} + {\bf 
\Omega_R (k)} \right ] \nonumber \\
& = & {\frac{2 m^* a_{42}}{e \hbar}} \left [ -\left ({\frac{m \pi}{W_z}} 
\right )^2 + \left ({\frac{n \pi}{W_y}} \right )^2 \right ] k_x \hat{x} +
{\frac{2 m^* a_{46}}{e \hbar}} E_y k_x \hat{z}
\end{eqnarray}

Thus, {\bf B ({\bf k})} lies in the 
x-z plane and subtends an angle $\theta$ with the wire axis (x-axis) given by
\begin{equation}
\theta = \text{arctan}\left[ \frac{a_{46}E_y}{a_{42}\left(\frac{m\pi}{W_z} + 
\frac{n\pi}{W_y} \right ) \left(\frac{n\pi}{W_y} - \frac{m\pi}{W_z} \right )} 
\right]
\end{equation}

Note from the above that in any given subband in a quantum wire, the {\it 
direction} 
of ${\bf B ({\bf k})}$ is fixed, irrespective of the magnitude of the wavevector 
$k_x$, since $\theta$ is independent of $k_x$.
As a result, there is no D-P 
relaxation 
in any given subband, even in the presence of scattering.

However, $\theta$ is different in different subbands because 
the 
Dresselhaus 
interaction is different in different subbands. Consequently, as electrons 
transition between 
subbands because of inter-subband scattering, the  angle $\theta$, and therefore 
the 
direction of the effective 
magnetic field ${\bf B ({\bf k})}$, changes. This causes D-P relaxation
in a {\it multi-subband} quantum wire. Since 
spins 
precess about different axes in different subbands, ensemble averaging 
over electrons in all subbands results in a gradual decay of the net spin 
polarization. Thus, there is no D-P spin relaxation in a quantum wire if 
a single subband is occupied, but it is present if multiple subbands are 
occupied and 
inter-subband scattering occurs. This was  shown rigorously in ref. 
\cite{pramanik_transnano}.

The subband structure is therefore critical to D-P spin relaxation in a quantum 
wire. In fact, if a situation arises whereby all electrons transition to a 
single subband and remain there, further spin relaxation due to the D-P 
mechanism will cease thereafter. In this case, spin no longer decays, let alone 
decay exponentially with distance. Hence, spin depolarization (or spin 
relaxation) cannot be parameterized by a constant spin diffusion length. 

The rest of this paper is organized as follows. In the next section, we describe 
our model 
system, followed 
by results and discussions in section IV. Finally, we
conclude in section V.

\section{Model of upstream spin transport}

\begin{figure}
\epsfig{file=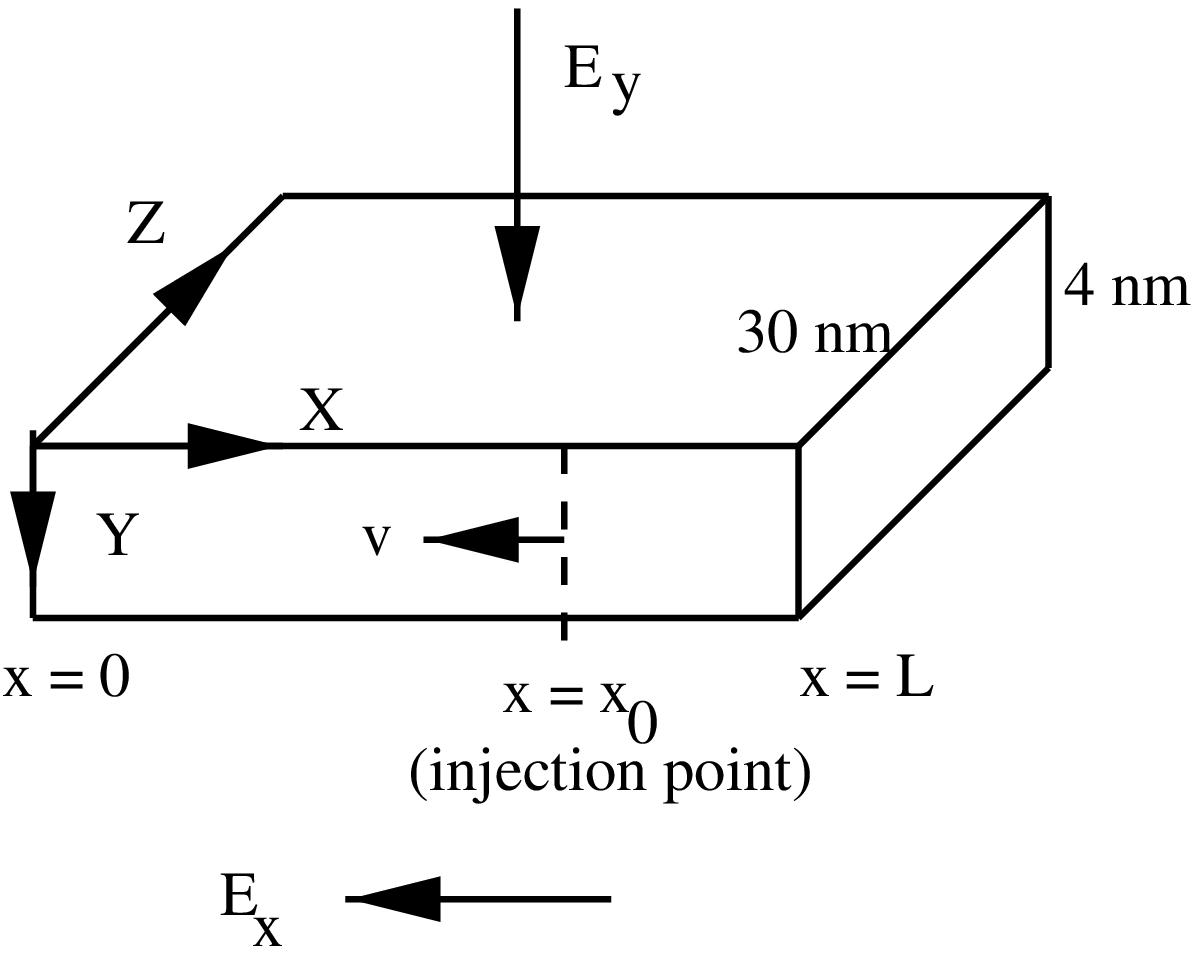, width=6in}
\caption{Pramanik et al.}
\label{model}
\end{figure}

\begin{figure}
\epsfig{file=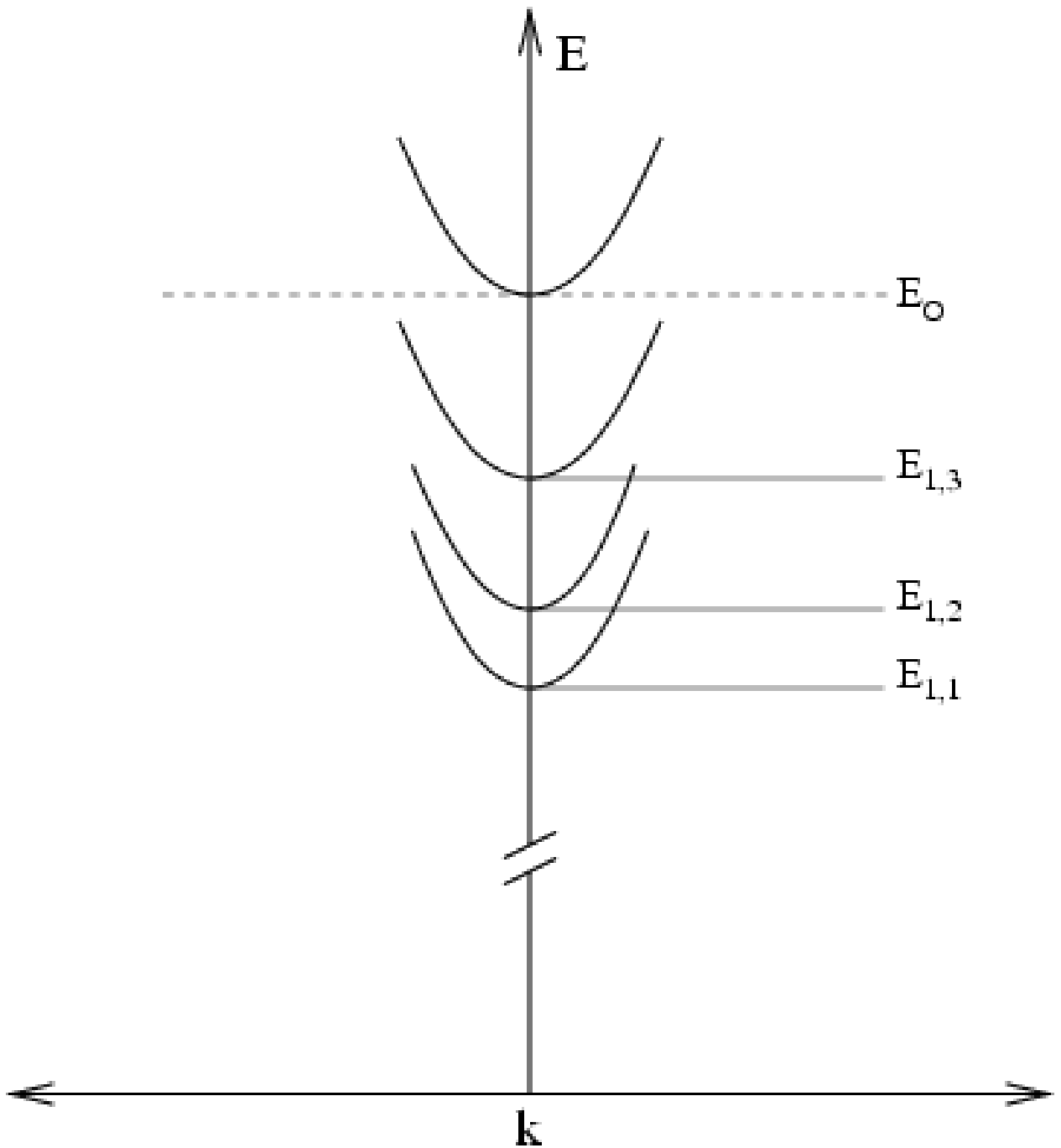, width=6in}
\caption{Pramanik et al.}
\label{subband}
\end{figure}

We consider a non centro-symmetric (e.g. GaAs) quantum wire with axis along 
[100] crystallographic direction. We choose a three dimensional Cartesian 
coordinate system with $\hat x$ coinciding with the axis of the quantum wire 
(refer Fig. \ref{model}). The structure is of length $L_x=1.005\
 \mu m$ with rectangular cross section: $W_y= 4$ nm and $W_z=30$ nm. A metal 
gate is placed on the top (not shown in Fig. \ref{model}) to
 induce the symmetry breaking
electric field $E_y\hat y$, which causes the Rashba interaction. In a quantum 
wire defined by split Schottky gates on a two-dimensional electron gas, $E_y\hat 
y$
arises naturally because of the triangular potential confining carriers near
the heterointerface.
We assume 
$E_y=100$ kV/cm \cite{pramanik_prb}. In addition, there
is another electric field $-E_x\hat x$  ($E_x>0$)
which drives transport along the axis of the quantum wire. Consider the case 
when {\it{spin polarized}} monochromatic electrons are constantly injected into 
the
 channel at $x=x_0=1\ \mu m$ with injection velocities along $-\hat x
$. If these electrons occupy only the lowest subband {\it at all times}, then 
there will be no D'yakonov-Perel' relaxation
\cite{pramanik_transnano}. Therefore, in order to study multisubband effect on 
spin dephasing of upstream electrons, we inject them with enough energy ($E_0$) 
that
 they initially occupy multiple subbands. We ignore any thermal broadening of 
injection energy \cite{saikin1} since $E_0>>k_BT$ for the range of temperature
 ($T$) considered, $k_B$ being Boltzmann constant. Let $E_i$ denote the energy 
at the bottom of $i$th subband ($i=1,2,\cdot\cdot\cdot\ n, n+1, 
\cdot\cdot\cdot$ etc.). We place $E_0$ between the $n$-th and $(n+1)$-th subband 
bottoms as shown in Fig. \ref{subband}. In other words,  $E_n<E_0<E_{n+1}$. We 
assume 
that the injected 
electrons, each with energy $E_0$, are distributed uniformly over the lowest $n$ 
subbands. In other wordes, at time $t=0$, electron population of the $
i$ th $(1\leq i\leq n)$ subband is given by $N_i(x,t=0)= (N_0/n)\ 
\delta(x-x_0)\ \delta(E-E_0)$ where $N_0$ is the total number of injected 
electrons and $E$ denotes their energies. At any subsequent time $t$, these
 distributions spread out in space ($x<x_0$), as well as in energy,  due to 
interaction of the injected 
electrons with the electric field $E_x$ and numerous  scattering events. 
Relative 
population of electrons among different subbands will change as well
 due to intersubband scattering events. {\it Upstream } electrons originally 
injected 
into, 
say, subband $i$ with velocity $-v_i\hat x$ ($v_i>0$), gradually 
slow down because of scattering and the decelerating electric field. They change 
their direction of motion (i.e. become {\it{downstream}}) 
beyond a distance $|\overline{x_i} - x_0|$ measured from the injection point 
$x_0$.
 Thus, no electrons will be found in the $i$-th 
subband beyond $\overline{x_i}$. Note that the value of $|\overline{x_i} - x_0|$
depends on three factors: the initial injection velocity into subband $i$, the 
decelerating electric
field and the scattering history. On the other hand, the `classical 
turning distance' of monochromatic electrons injected into the $i$-th subband 
with energy $
E_0$ for a given electric field $E_x$ is given by
\begin{equation}
E_0 - E_i = (1/2)m^* v_i^2 = e E_x |x_i - x_0|
\label{clas}
\end{equation}
where $E_i$ is the energy at the bottom of the $i$-th subband and $v_i$ is the 
injection velocity in the $i$-th subband. Note that $x_i$ does not depend on 
scattering 
history and $x_i$ = $\overline{x_i}$ in ballistic transport.

Clearly $v_n< 
v_{n-1}<\cdot\cdot\cdot<v_2<v_1$ for a given $E_0$ (see Fig. \ref{subband}). 
Thus, for a 
given channel 
electric field 
$E_x$, $x_n-x_0=$min$\{x_i-x_0\}, i=1, 2, ..., n$. Hence we concentrate on the 
region $(x_n, x_0)$ where, almost  {\it{all}} injected electrons are 
{\it{upstream}} electrons. In the simulation, velocity of every 
electron is tracked and as soon as an electron alters direction and goes 
downstream (i.e. its 
velocity becomes positive) it is ignored by the simulator and another upstream 
electron is simultaneously injected from $x=x_0$ randomly 
in any of the $n$ lowest subbands with equal probability. This process is 
continued for a 
sufficiently long time till electron distributions over different subbands, 
$N_i(x,t), i=1,2,...n$, no longer change with time. Under this condition we say 
that {\it{steady state}} is achieved for the upstream electrons. This steady 
state electron distribution is extended from $x=x_n$ to $x=x_0$ and heavily 
skewed near the region $x=x_0$. This steady state 
distribution of upstream electrons {\it does not} represent the local 
equilibrium 
electron distribution because of two reasons: (a) upstream electrons are 
constantly injected into the channel at $x=x_0$; this is the reason why the 
distribution is skewed near $x=x_0$ and (b) we exclude any 
downstream electron from the distribution. At local equilibrium, there will be 
of course both upstream and downstream electrons in the distribution.\par

The model above allows us to separate upstream electrons from downstream 
electrons and therefore permits us to study upstream electrons in isolation.
Of course, in a real quantum wire, both upstream and downstream electrons will 
be present at any time, even in the presence of a strong electric field, since 
there will be always some non-vanishing contribution of back-scattered electrons 
to the upstream population.

The semiclassical model and the simulator used to simulate spin transport have 
been described in ref. \cite{pramanik_prb}. Based
on that model, at steady state, the magnitude of the ensemble averaged spin 
vector at any position $x$ inside the channel is given by
\begin{equation}
\left|\langle{\bf{S}}\rangle\right|(x)=\frac{\sqrt{\left(\underset{i=1}{\overset{n}{\sum}}N_i(x)
\left\langle S_{ix}\right\rangle(x)\right)^2+\left(\underset{i=1}{\overset{n}{\sum}}N_i(x)\left\langle S_{iy}\right\rangle(x)\right)^2+\left(\underset{i=1}{\overset{n}{\sum}}N_i(x)\left\langle S_{iz}\right\rangle(x)\right)^2}}{\underset{i=1}{\overset{n}{\sum}}N_i(x)} 
\label{avgspin}
\end{equation}

Here $\left\langle S_{i\zeta}\right\rangle(x)$, $\zeta= x, y, z$, denotes the 
ensemble average of $\zeta$ component of spin at position $x$. Subscript $i$ 
implies that ensemble averaging is carried out over electrons {\it{only}} in the 
$i$ 
th subband. The above equation can be simplified to
\begin{equation}
\left|\langle{\bf{S}}\rangle\right|(x)=\frac{\sqrt{\left(\underset{i=1}{\overset{n}{\sum}}N_i(x)\left|\langle{\bf{S}}_i\rangle(x)\right|\right)^2-2\underset{i,j=1}{\overset{n}{\sum}}N_i(x)N_j(x)\left|\langle{\bf{S}}_i\rangle(x)\right|\left|\langle{\bf{S}}_j\rangle(x)\right|\sin^2\frac{\theta_{ij}(x)}
{2}}}{\underset{i=1}{\overset{n}{\sum}}N_i(x)}
\label{avgspin1}
\end{equation}
where 
$\left\langle{\bf{S}}_l\right\rangle(x)=\left\langle{{S}}_{lx}\right\rangle(x
)\hat x + \left\langle{{S}}_{ly}\right\rangle(x)\hat y + 
\left\langle{{S}}_{lz}\right\rangle(x)\hat z$ and $\theta_{ij}(x)$ is the 
angle between $\left\langle{\bf{S}}_i\right\rangle(x)$ and 
$\left\langle{\bf{S}}_j\right\rangle(x)$.
Note that in absence of any intersubband scattering event, 
$\left|\langle{\bf{S}}_l\rangle\right|(x)=1$ for all $x$ (i.e. initial spin 
polarization of the injected electrons) \cite{pramanik_transnano}. Simulation 
results that we present in 
the next section can be understood using Equation (\ref{avgspin1}). 

\section{Results and Discussion}
We examine how ensemble averaged spin polarization of upstream electrons 
$\left|\langle{\bf{S}}\rangle\right|(x)$ varies in space for different values of 
driving electric 
field $E_x$ and injection energy ($E_0$) for a fixed lattice temperature $T$. We 
vary $E_x$ in the range $0.5-2$kV/cm for constant injection energy $426$ meV and 
$T$ = 30 K, where $E_0$ is measured from the bulk conduction band energy as 
shown in Fig. \ref{subband}. The lowest subband bottom is 351 meV above the bulk 
conduction band edge. We also present results corresponding to 
$E_0=441$ meV with $E_x=1$kV/cm and $T = 30$K. In all cases mentioned above, 
injection energies lie between subband 3 and subband 4. Injected electrons are 
equally distributed among the three lowest subbands initially. Obviously, this 
corresponds to a non-equilibrium situation.  All injected electrons are 100\% 
spin polarized transverse to the wire axis (i.e either $\hat y$ or $\hat 
z$).\par

\begin{figure}
\epsfig{file=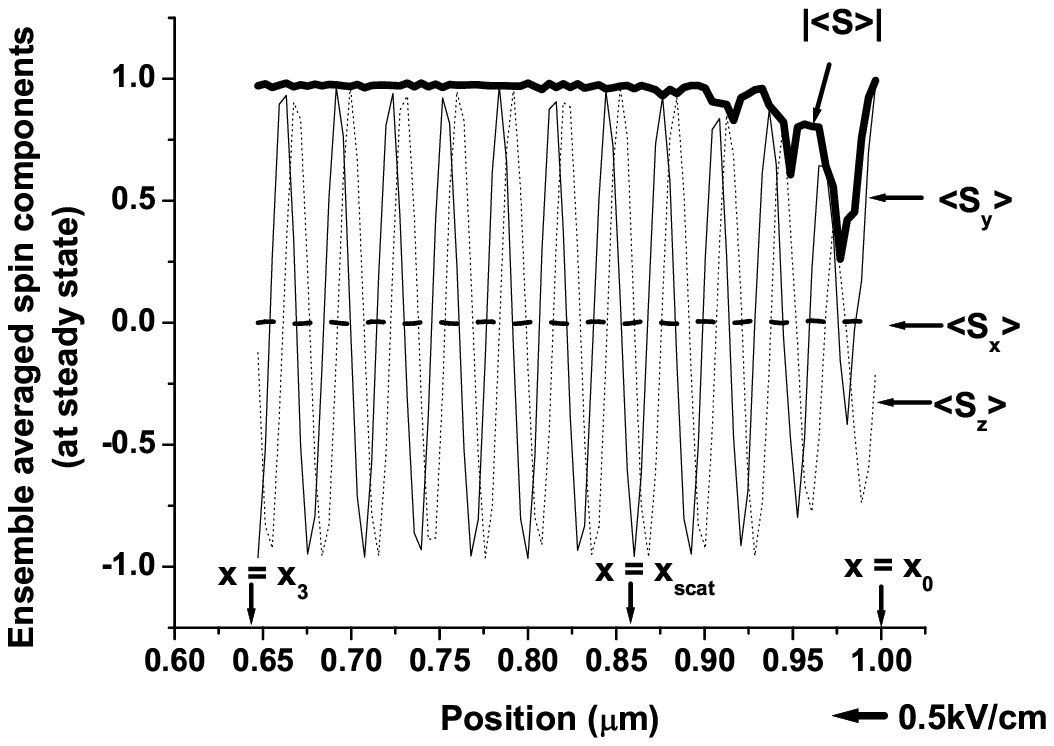, width=6in}
\caption{Pramanik et al.}
\label{halffieldspin}
\end{figure}

\begin{figure}
\epsfig{file=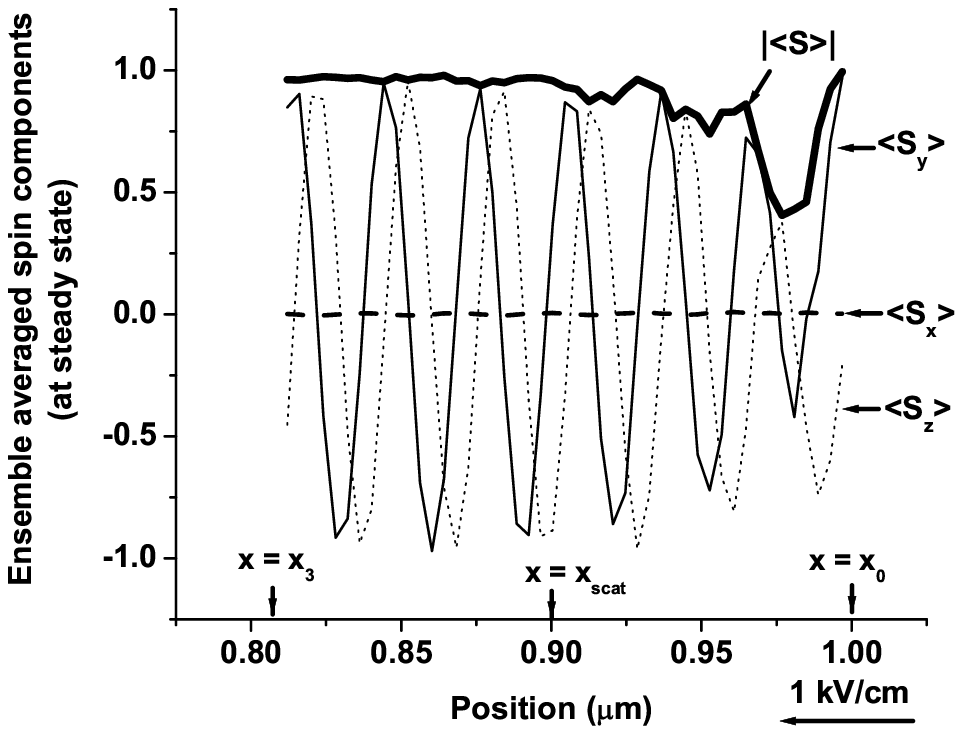, width=6in}
\caption{Pramanik et al.}
\label{onefieldspin}
\end{figure}

\begin{figure}
\epsfig{file=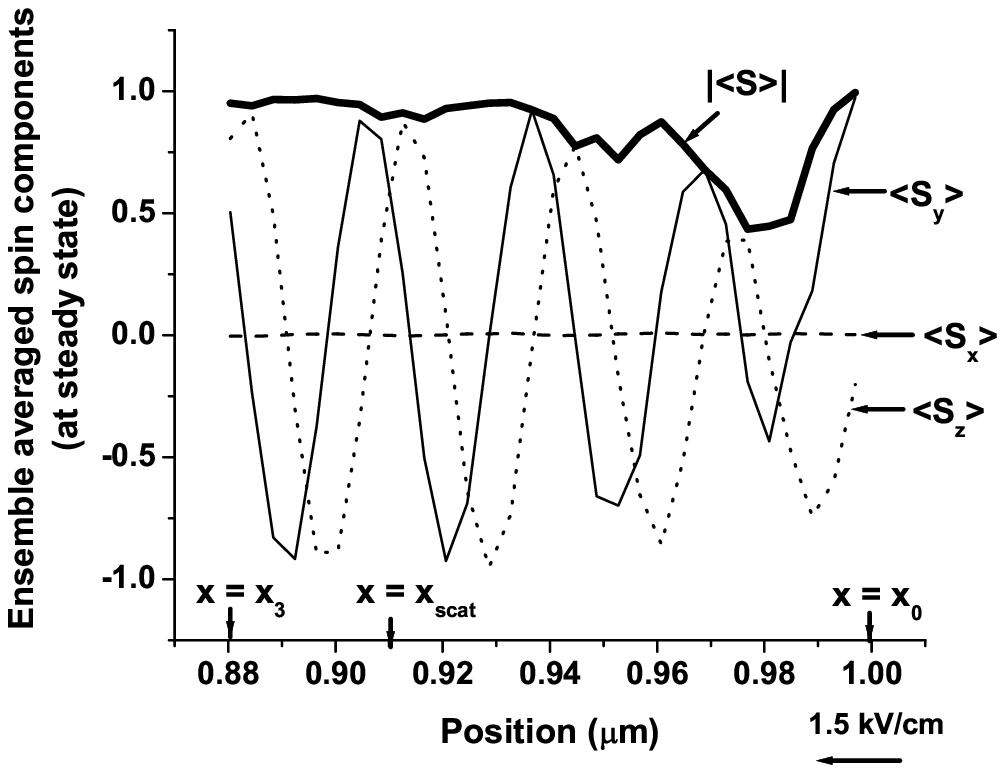, width=6in}
\caption{Pramanik et al.}
\label{oneandhalffieldspin}
\end{figure}

\begin{figure}
\epsfig{file=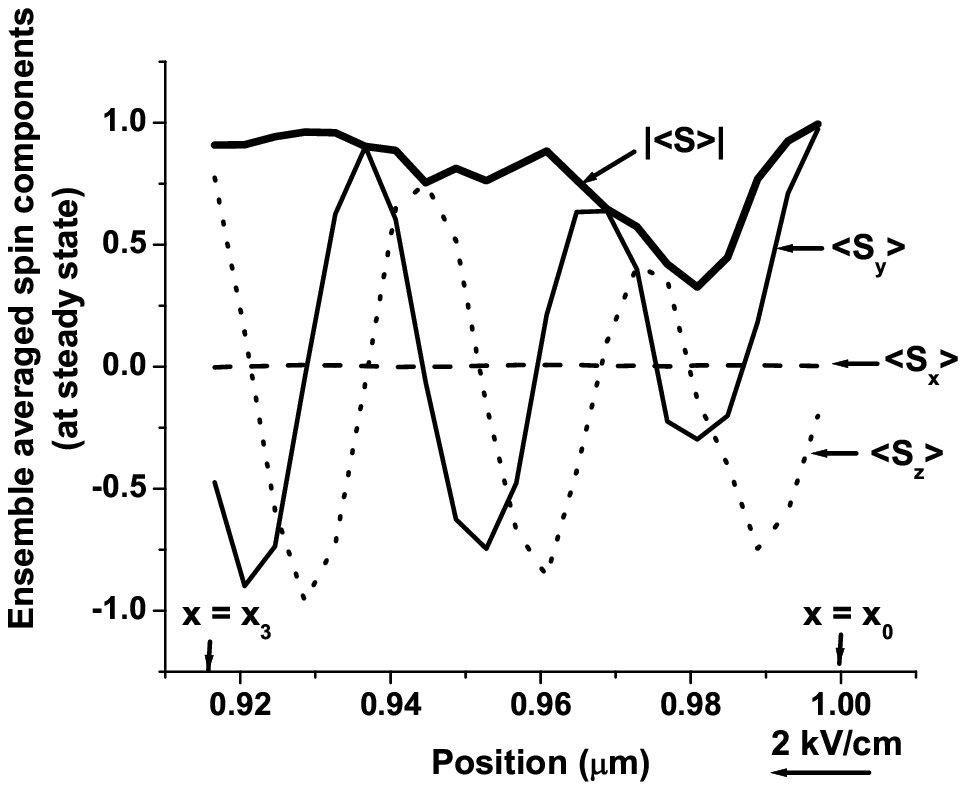, width=6in}
\caption{Pramanik et al.}
\label{twofieldspin}
\end{figure}

\begin{figure}
\epsfig{file=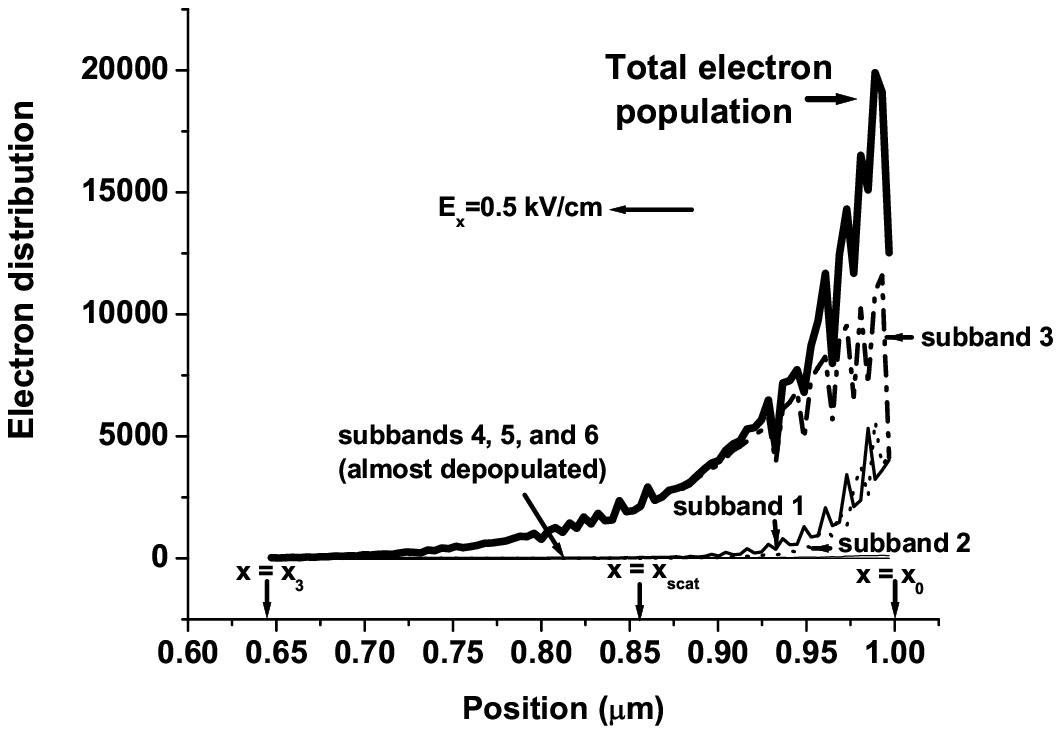, width=6in}
\caption{Pramanik et al.}
\label{halffieldsub}
\end{figure}

\begin{figure}
\epsfig{file=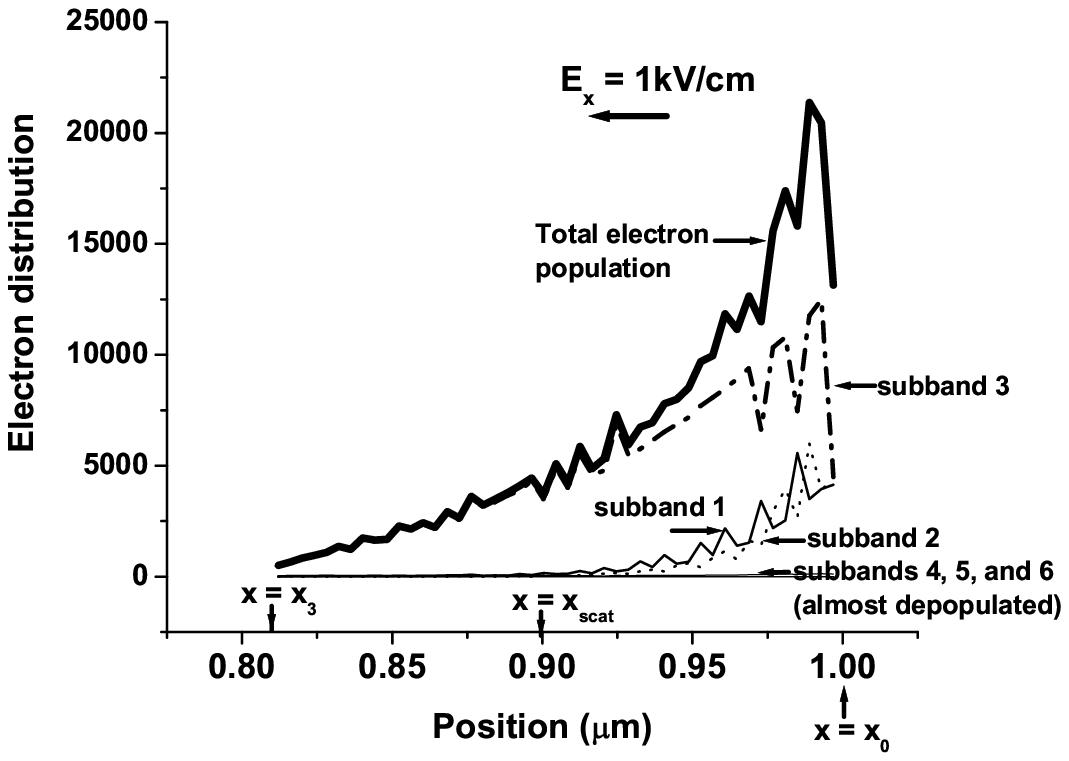, width=6in}
\caption{Pramanik et al.}
\label{onefieldsub}
\end{figure}

\begin{figure}
\epsfig{file=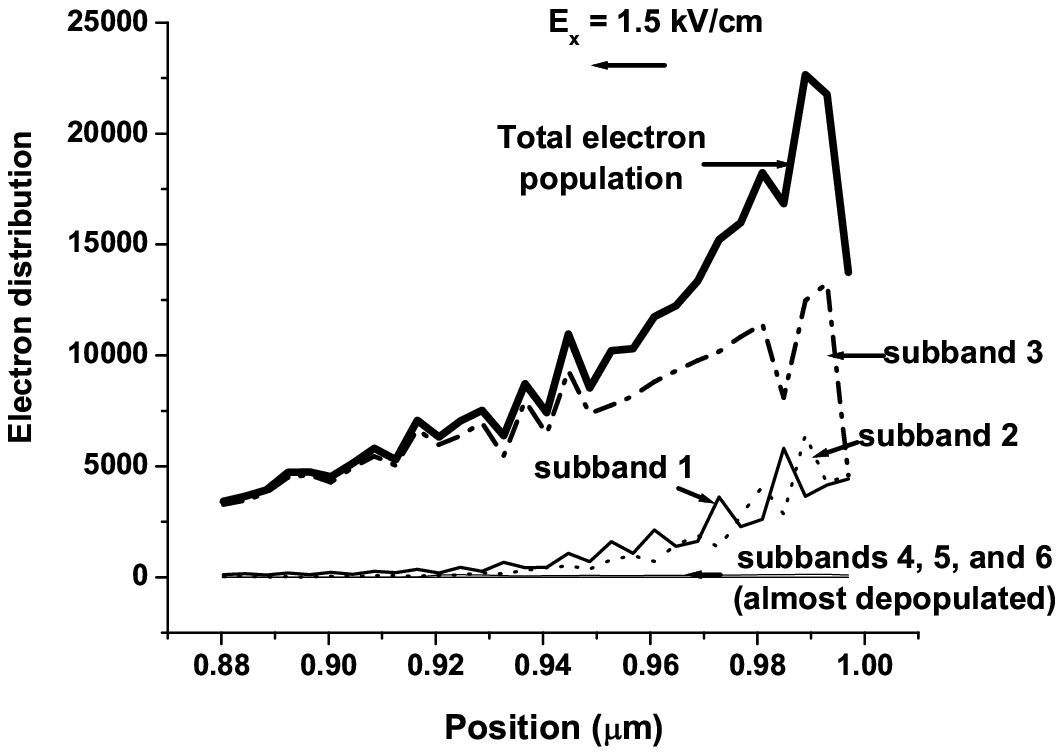, width=6in}
\caption{Pramanik et al.}
\label{oneandhalffieldsub}
\end{figure}

\begin{figure}
\epsfig{file=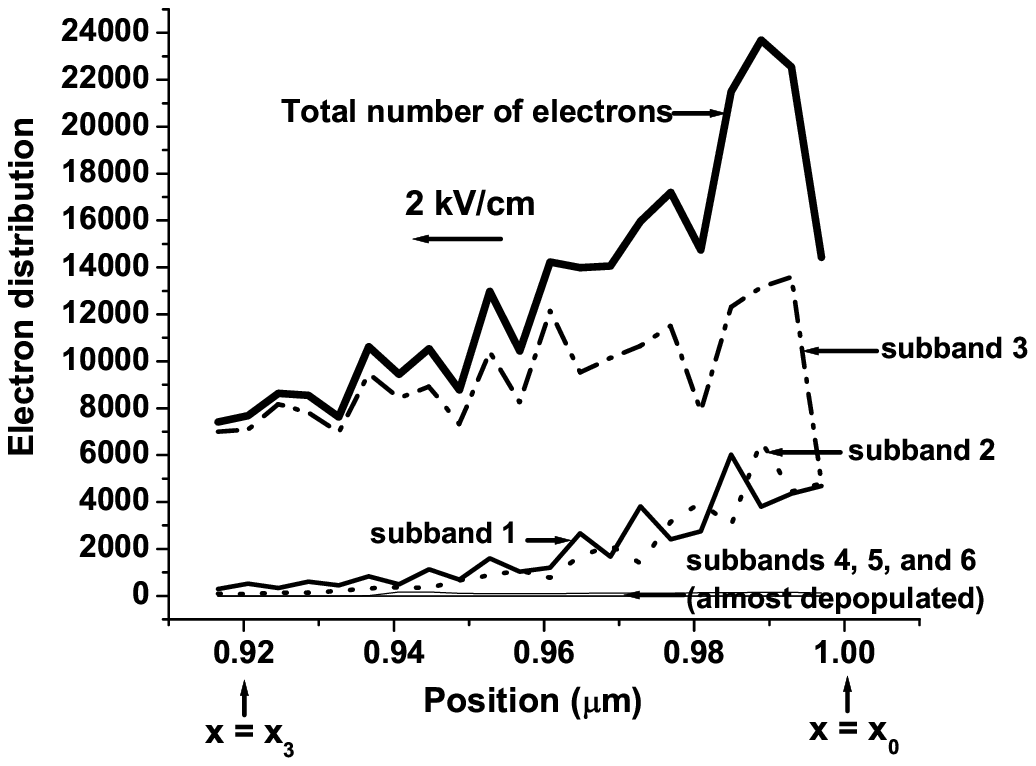, width=6in}
\caption{Pramanik et al.}
\label{twofieldsub}
\end{figure}

\begin{figure}
\epsfig{file=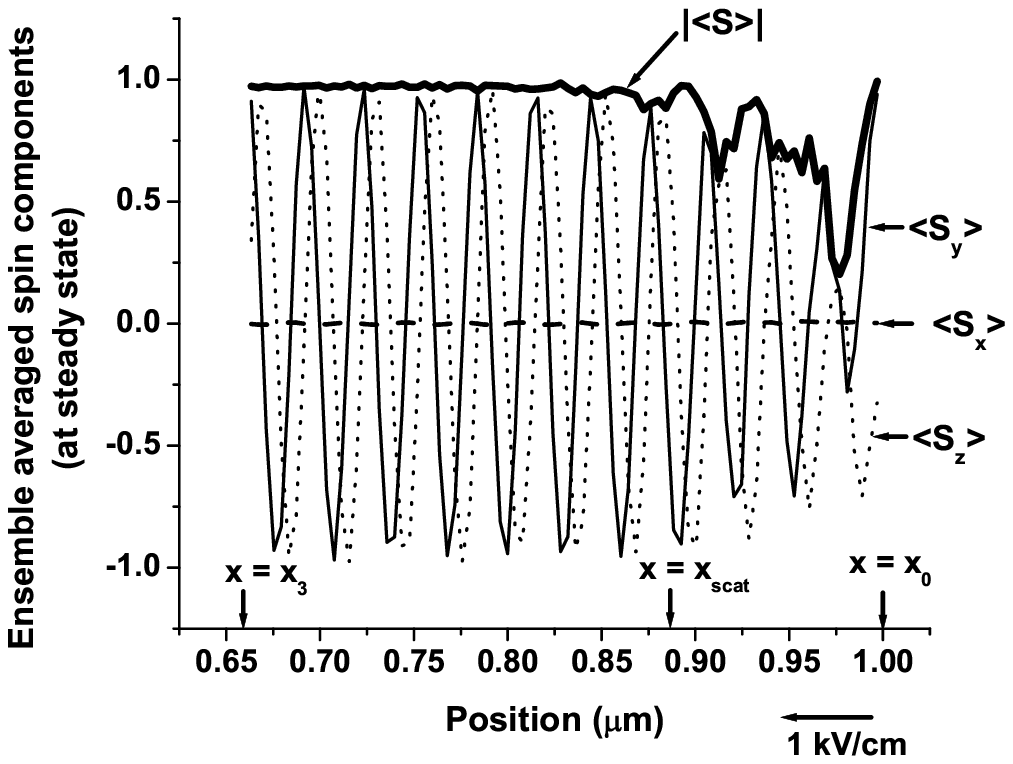, width=6in}
\caption{Pramanik et al.}
\label{90mev}
\end{figure}

\begin{figure}
\epsfig{file=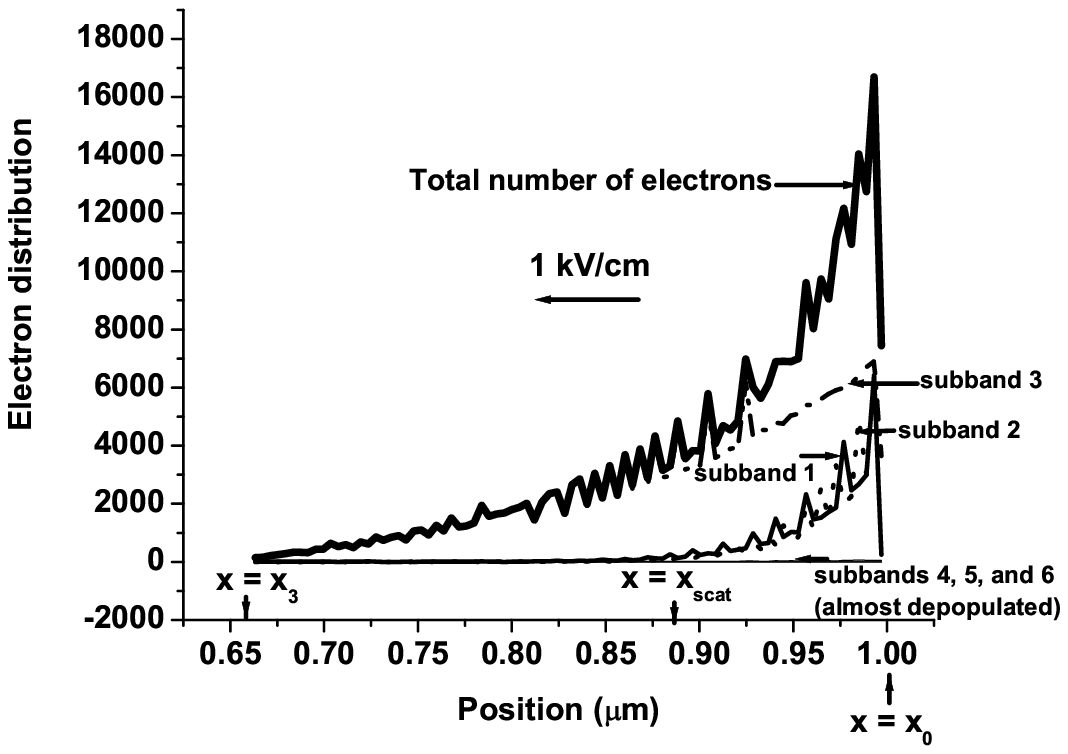, width=6in}
\caption{Pramanik et al.}
\label{sub90mev}
\end{figure}

\begin{figure}
\epsfig{file=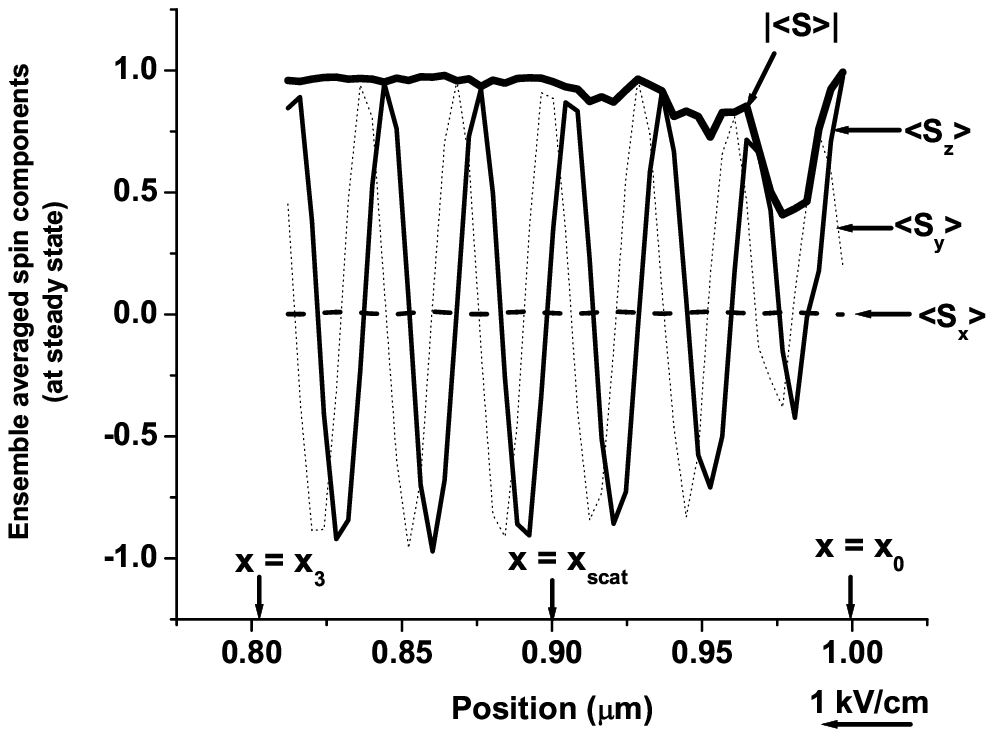, width=6in}
\caption{Pramanik et al.}
\label{pos75mev-z}
\end{figure}

Figures \ref{halffieldspin}-\ref{twofieldspin}, \ref{90mev}, and 
\ref{pos75mev-z} show how ensemble averaged spin components 
$\langle S_x\rangle(x)$, 
$\langle S_y\rangle(x)$, $\langle S_z\rangle(x)$ and $|\langle{\bf 
S}\rangle|(x)$ of upstream electrons evolve over 
space. Figures \ref{halffieldspin}-\ref{twofieldspin} show the influence of the 
driving electric
field on spin relaxation, Fig. \ref{90mev} shows the influence of initial 
injection energy and
Fig. \ref{pos75mev-z} shows the influence of the intial spin polarization. It
is evident that neither the driving electric field, nor the initial injection 
energy, 
nor the intial spin polarization has any significant effect on spin relaxation.
Note that $|\langle{\bf S}\rangle|(x)$ does {\it not} decay exponentially 
with distance, 
contrary to Equation (\ref{decay}). Spatial distribution of electrons over 
different subbands is shown in Fig. \ref{halffieldsub} - 
\ref{twofieldsub}, and Fig. \ref{sub90mev}. The classical turning 
point of  electrons in the third subband ($x_3$) has been indicated in each 
case. 
Fig. \ref{halffieldsub} - 
\ref{twofieldsub} show the influence of the driving electric field and  Fig. 
\ref{sub90mev} shows the 
influence of initial injection energy on the spatial evolution of subband 
population.
As expected, $|x_0 - x_3|$ decreases with increasing electric field in 
accordance
with Equation (\ref{clas}). Note that at low electric field (Figs. 7 and 8) 
$\overline{x_3} \approx x_3$ since all subbands are getting nearly depopulated 
of 
``upstream'' electrons 
at $x$ = $x_3$. Recall that $\overline{x_3} = x_3$ only if transport is 
ballistic; 
therefore 
we can 
conclude 
that upstream transport is 
nearly ballistic in the range $|x_0 - x_3|$ when $E_x$ $<$ 1 kV/cm. 
At high electric field, when $E_x$ $>$ 1.5 kV/cm, (Fig. 10) $|\overline{x_3} - 
x_0| > |x_3 - x_0|$.  
This indicates that there are many upstream electrons even beyond the classical
turning point. It can only happen if there is significant scattering 
that drives electrons against the electric field, making them go beyond the 
classical turning point.
We can also deduce that most of these scattering events imparts momentum to the 
carriers to {\it aid}
 upstream motion rather than oppose it, since $|\overline{x_3} - x_0| > |x_3 - 
x_0|$. This behavior 
is a consequence of the precise nature of the scattering events and would not 
have
been accessible in drift-diffusion models that typically treat scattering 
via a relaxation time approximation. 

\subsection{Population inversion of upstream electrons}

 Note that  even though electrons are injected equally into all three subbands, 
most electrons end up in subband 3 -- the highest subband occupied initially -- 
soon after 
injection. Beyond a certain distance ($x=x_{scat}\approx 0.9\ \mu m$) subbands 1 
and 2 become virtually depopulated. This feature is very counter-intuitive and 
represents a {\it population inversion} of upstream electrons! It can be 
understood as follows: 
scattering rate of an electron with energy $E$ is proportional to the density of 
the final state. In a quantum wire, density of states has $1/\sqrt{E-E_i}$ 
dependence where $E_i$ is the energy at the bottom of the $i$th subband. As the 
injected electrons move upstream they gradually cool down and their energies 
approach the energy at the bottom of subband 3 ($E_3$). To visualize this, 
imagine the horizontal line $E_0$ in Fig. \ref{subband}  sliding down with 
passage of 
time. As $E_3$ is approached, electrons will increasingly scatter into subband 3 
since the density of final state in subband 3 is increasing rapidly. To scatter 
into a final state in subband 2 or 1 that has the same density of state as in 
subband 3 will  require a much larger change in energy and hence a much more 
energetic phonon which is rare since the phonons obey Bose Einstein statistics.
Therefore, subband 3 is the overwhelmingly preferred destination and this 
preference increases rapidly as electrons cool further. Consequently, 
beyond a certain distance, virtually all electrons are scattered to subband 3 
leaving subbands 1 and 2 depleted. 
This feature is a peculiarity of quasi one-dimensional system and will {\it not} 
be observed in bulk or quantum wells.
Exact values of $x_3$ and $x_{scat}$ depend on injection energy and electric 
field. In the field range $0.5-1.5$kV/cm and injection energy $426$ meV, $|x_3| 
> |x_{scat}|$. However, for higher values of electric field (e.g.\ 2kV/cm) or 
smaller values of injection energies, electrons reach classical turning point 
even before subbands 1 and 2 get depopulated. \par

Because of electron bunching in subband 3, spin dephasing in the region 
$(x_{scat},x_0)$ is governed by Equation (6)  with 
$n=3$. We observe a few subdued oscillations in 
$\left|\langle{\bf{S}}\rangle\right|(x)$ in 
this region because of the ``sine term'' in Equation (\ref{avgspin1}). However, 
in the region $(x_3,x_{scat})$ subbands 1 and 2 are almost depopulated. 
Therefore, there is no D-P relaxation in the interval $(x_3,x_{scat})$ since 
only a single subband is occupied \cite{pramanik_transnano}. Consequently, the 
ensemble averaged spin assumes a constant value 
$\left|\langle{\bf{S}}_3\rangle\right|<1$ and does not change any more.  Thus in 
this region, one can say that spin dephasing 
length becomes infinite. It should be noted that it is meaningless to study spin 
dephasing in the region $x<x_3$ because electrons  do not even reach this 
region.\par

\section{Conclusion}
In this paper, we have used a semiclassical model to study spin dephasing of 
upstream electrons in a quantum wire, taking into account the subband formation. 
We showed that the subband structure gives rise to 
rich features in the spin dephasing characteristics of upstream electrons that 
cannot be captured in models which fail to account for the precise physics of 
spin dephasing and the fact that it is different in different subbands.
Because spin relaxation in a multi-subband quantum wire is non-exponential (even 
non-monotonic) in space, it does not make sense to invoke a ``spin 
diffusion length'', let alone use such a heuristic parameter to model spin 
dephasing.

Finally, we have found a population inversion effect for upstream electrons. It 
is possible that downstream electrons also experience a similar population 
inversion. This scenario is currently being investigated.

\bigskip

\noindent {\bf Acknowledgement}: The work at Virginia Commonwealth University
is supported by the Air Force Office of Scientific Research under grant
FA9550-04-1-0261.

\clearpage
{\bf{Figure captions :}}\\ \\

Figure 1. A quantum wire structure of length $L=1.005\mu m$ with rectangular 
cross section 30 nm $\times$ 4 nm. A top gate (not shown) applies a symmetry 
breaking electric field 
$E_y$ to induce the Rashba interaction. A battery (not shown) applies an 
electric 
field $-E_x\hat x$  ($E_x>0$), along the channel. Monochromatic spin polarized 
electrons are 
injected at $x=x_0=1\mu m$ with injection velocity $-v_{inj}$. These electrons 
travel along $-\hat x$ ({\it{upstream}} electrons) until their direction of 
motion is reversed due to the electric field $-E_x\hat x$. We investigate spin 
dephasing of these upstream electrons.\\

Figure 2. Subband energy dispersion in the quantum wire.\\

Figure 3. Spatial variation of ensemble averaged spin components for driving 
electric field $E_x$ = 0.5kV/cm at steady state. 
Lattice temperature is $30$ K, injection energy $E_0 = 426$ meV. Electrons are 
injected with equal probability into the three lowest subbands. Classical 
turning point of subband $3$ electrons is denoted by $x_3$ and $x_{scat}$ 
indicates the point along the channel axis where subbands $1$ and $2$ gets 
virtually depopulated. Injected electrons are $\hat y$ polarized and $x=x_0=1\ 
\mu m$ is the point of injection.\\

Figure 4. Spatial variation of ensemble averaged spin components for driving 
electric field $E_x$ = 1kV/cm at steady state. 
 Other conditions are same as in Figure 3.\\

Figure 5. Spatial variation of ensemble averaged spin components for driving 
electric field $E_x$ = 1.5kV/cm at steady state.  Other conditions are same as 
in Figure 3.\\

Figure 6. Spatial variation of ensemble averaged spin components for driving 
electric field $E_x$ = 2kV/cm at steady state. Other conditions are same as in 
Figure 3.\\

Figure 7. Spatial variation of electron population over different subbands at 
steady state for driving electric field $E_x$ = 0.5 kV/cm. 
Other conditions are same as before.\\

Figure 8. Spatial variation of electron population over different subbands 
at steady state for driving electric field $E_x$ = 1 kV/cm. 
Other conditions are same as before.\\

Figure 9. Spatial variation of electron population over different subbands 
at steady state for driving electric field $E_x$ = 1.5 kV/cm. 
Other conditions are same as before.\\

Figure 10. Spatial variation of electron population in different subbands 
at steady state for driving electric field $E_x$ = 2 kV/cm. 
Other conditions are same as before.\\

Figure 11. Spatial variation of ensemble averaged spin components for $E_0 = 
441$meV, $E_x=1$kV/cm and lattice temperature $T = 30$K. Injected electrons 
are $\hat y$ polarized.\\

Figure 12. Spatial variation of electron population over different subbands at 
steady state  for $E_0 = 441$meV, $E_x=1$kV/cm and lattice temperature $T = 
30$K. Injected electrons are $\hat y$ polarized.\\

Figure 13. Spatial variation of ensemble averaged spin components for $E_0 = 
426$meV, $E_x=1$kV/cm and lattice temperature $T = 30$K. Injected electrons are 
$\hat z$ polarized.\\

\clearpage
%%%%%%%%%%%%%%%%%%%%%%%%%%%%%%%%%%%%%%%%%%%%%%%%%%%%%%%%%%%%%%%%%%%%%%%%%%%%%%%%%%%%%%%

\end{document}